# Semantic Inference using Chemogenomics Data for Drug Discovery

Qian Zhu (qianzhu@indiana.edu)
Yuyin Sun (yuysun@indiana.edu)
Sashikiran Challa (schalla@umail.iu.edu)
Ying Ding (dingying@indiana.edu)
Michael S Lajiness (LAJINESS_MICHAEL_S@lilly.com)
David J Wild (djwild@indiana.edu)

# Semantic Inference using Chemogenomics Data for Drug Discovery

Qian Zhu*[§1], Yuyin Sun*[2], Sashikiran Challa*[1], Ying Ding*[2], Michael S Lajiness*[3], David J Wild*[1]

[1] School of Informatics and Computing, Indiana University, Bloomington, IN 47408, USA

[2] School of Library and Information Science, Indiana University, Bloomington, IN 47408, USA

[3] Eli Lilly and Company, Indianapolis, IN 46225, USA

*These authors contributed equally to this work

[§]Corresponding author

Email addresses:

    QZ: qianzhu@indiana.edu

    YYS: yuysun@indiana.edu

    SC: schalla@umail.iu.edu

    DY: dingying@indiana.edu

    MSL: LAJINESS_MICHAEL_S@lilly.com

    DJW: djwild@indiana.edu

# Abstract

**Background**

Semantic Web Technology (SWT) makes it possible to integrate and search the large volume of life science datasets in the public domain, as demonstrated by well-known linked data projects such as LODD, Bio2RDF, and Chem2Bio2RDF. Integration of these sets creates large networks of information. We have previously described a tool called WENDI for aggregating information pertaining to new chemical compounds, effectively creating evidence paths relating the compounds to genes, diseases and so on. In this paper we examine the utility of automatically inferring new compound-disease associations (and thus new links in the network) based on semantically marked-up versions of these evidence paths, rule-sets and inference engines.

**Results**

Through the implementation of a semantic inference algorithm, rule set, Semantic Web methods (RDF, OWL and SPARQL) and new interfaces, we have created a new tool called Chemogenomic Explorer that uses networks of ontologically annotated RDF statements along with deductive reasoning tools to infer new associations between the query structure and genes and diseases from WENDI results. The tool then permits interactive clustering and filtering of these evidence paths.

**Conclusions**

We present a new aggregate approach to inferring links between chemical compounds and diseases using semantic inference. This approach allows multiple evidence paths between compounds and diseases to be identified using a rule-set and semantically annotated data, and for these evidence paths to be clustered to show overall evidence linking the compound to a disease. We believe this is a powerful approach, because it allows compound-disease relationships to be ranked by the amount of evidence supporting them.

## Background

Recent advances in chemical & biological sciences have led to an incredible increase in the volume of information about known chemical compounds, genes, diseases, and assays. Statistical data from the PubChem Substance Database of chemical structures, shows an increase from 35,379,748 structures in 2007 to 69,088,100 in 2010; the number of PubChem Bioassays increased from around 1000 in 2008 to 434,635 in 2010[1], and there are 726,872 compound records and 2,925,588 activities in the chemogenomic ChEMBL[2] dataset. Numerous other chemical, chemogenomic, and biological data (including data extracted from the scholarly literature) are also available including ChEBI[3], CTD[4], KEGG[5] and Medline[6] inter alia. Many well-known search engines for these data resources have been developed like PubChem, which provides chemical structure search and bioassay search. This search engine returns an abundant supply of chemical information and bioactive information based on PubChem Bioassay data. ChemSpider[7] links together compound information across the Web and provides free text and structure search access to millions of chemical structures. It offers multiple search modes to do chemical information searching on the basis of hundreds of data vendors.

We can imagine all these information resources as buckets for pieces of a very large jigsaw puzzle, each bucket containing only pieces of a certain color. To assemble the full picture we need to be able to search and apply algorithms that span across different buckets seamlessly. There are many technologies of utility for this, most recently from the Semantic Web Technology (SWT) community, like XML (for describing data), OWL (for describing ontologies and taxonomies), and RDF (for describing relationships) are allowing data aggregation and the representation of meaning and relationships in the data, and are now being quite widely applied for life science data. LODD[8], Bio2RDF[9], Chem2Bio2RDF[10] demonstrate not only how SWT can enable integration of multiple sources, but also complex query processing using the SPARQL language. Our resource, Chem2Bio2RDF, integrates six categories of data based on the kinds of biological and chemical concepts and relationships they represent: chemical & drug, protein & gene, chemogenomics, systems (i.e., PPI and pathway), phenotype (i.e., disease and side effect) and literature. However, the current version of Chem2Bio2RDF lacks a formal ontology so it is hard for users to read and understand the meaning of the metadata and harder to do further inference. Once an integrated network of compounds, genes, diseases, etc. is in place (with an appropriate ontology), as PharmGKB[11] establishes knowledge about the relationships among drugs, diseases and genes, including their variations and gene products, it becomes possible to semantically infer new links in the network (i.e. identify new associations) via sets of rules, and inference engines that use these rules. For example, we might have a rule that if a chemical compound A is highly similar to a drug D that is known to be active against a protein target T, we infer an association (and thus a network link) between A and T (possibly annotated with a confidence value). Semantic inference has been used in various applications including knowledge-based recommender systems[12] and human-machine communication[13], but there have few applications in the life sciences, Neurocommons[14] uses SWT for assembling and querying biomedical knowledge from multiple sources and disciplines. With this system, scientists will be able to load in lists of genes that come off the lab robots, and get back those lists of genes with relevant information around them based on the public knowledge[15]. SciNetS

Search[16] is an inference search over integrated life science databases using SWT. It can implement cross-domain search and use statistical scoring. All the metadata of databases are described as a set of triples consisting of two bio-items and relationships between these items. GoRouter[17] is building an RDF model to do semantic query and inference, but the inference is restricted to the Gene Ontology and its related associations.

In our previous paper [18], we introduced a novel tool, WENDI (Web Engine for Non-Obvious Drug Information), for aggregating information related to a compound to identify relationships. WENDI probes the potential biological properties of the compound using predictive models, databases, and the scholarly literature, in particular, to find non-obvious relationships between the compound and assays, genes, and diseases, which cross over different types of data sources. The purpose of WENDI is not just to return data about a compound (such as in a database search): rather it allows a researcher to understand the context in which a compound operates, and to find clues which help them understand properties of the compound that they might not otherwise have discovered. WENDI does data integration for particular query compounds and represents its result graph in XML. WENDI architecture is shown in Figure 1.

**Figure 1. WENDI Architecture** (WENDI main web interface is show at the upper right corner)

WENDI has good performance on data integration, but it relies on the user manually find associations among the kinds of results presented. At this point, we thus extended WENDI work to use semantic inference and rules to automatically infer new associations based on the WENDI XML results. These new associations in aggregate form clusters of association that build evidence of an association between compounds and diseases via multiple sources or evidence paths. We have implemented this in a tool called Chemogenomic Explorer that uses networks of ontologically-enabled RDF statements (e.g. the query compound C is similar to compound D, drug D is active in assay A, assay A is associated with gene G) along with deductive reasoning tools to infer relationships between the query compound and genes and diseases, this will allow us to cluster insights by disease, and then prioritize the output based on the amount of evidence linking a compound to a disease.

## Methods

The WENDI web service is used to create an initial set of relational paths in XML. CE adds to the previously reported capabilities of WENDI through (i) the application and inference engine and rule set to enable new associations to be inferred; (ii) clustering and filtering of inferred evidence paths in a completely new interface and (iii) the application of Semantic Web languages and methods (RDF, SPARQL, OWL) to enable a much broader range of capabilities including creation and mining of evidence paths, and the annotation of relationships using the ontology. These new methods are described below.

WENDI XML includes the direct relationships between similar compounds and bioassays, similar compounds and literature references, bioassays and genes/diseases, and so on. The process of importing this information into CE is as follows: 1. Data preparation as described in section 2.1; 2. Semantic representation using a CE ontology and presentation in RDF format, described in section 2.2; 3. Rule-based Inference described in 2.3; and 4. Path ranking based on the number of properties for each disease described in 2.4.

**Data preparation**

WENDI aggregates information from diverse data sources and predictive models including PubChem Compound, PubChem Bioassay, PubChem3D [19], DrugBank[20], MRTD[21], CTD[4], ChEMBL[3], HuGEpedia[22], KEGG[5], and Medline[6]. Because not all of these sources have gene/disease terms related, we first extract the data with gene/disease information, such as PubChem Bioassay, CTD, ChEMBL, HuGEpedia and Medline. We employed different approaches according to the different datasets: for CTD, which already has compound-disease relation information, we extract such relationships directly; but for other data, the links between compounds and diseases are indirect. There are two ways to mine this information in the data preparation section. PubChem bioassay as an example, (i) implementing a SQL function "position" to find gene or disease terms from Phenopred Matrix[23] occurring in the description of the bioassay, then again based on the Phenopred Matrix to find associations between gene and disease, finally the link between bioassay-gene-disease can be established; (ii) using the GO ontology[24], we performed the same SQL clause to find which GO terms are noted in the description of bioassay, identified the genes associated with the GO term on the basis of GO annotation, then used the Phenopred matrix to find which diseases are linked to these genes. More information about this extraction can be found in our WENDI paper [18].

We extracted the above information from WENDI XML using XML DOM [25]. All the information is extracted into 4 groups: Active-Bioassay, CTD, Chembl, and Literature, which include compound, gene, disease, or bioassay and journal information.

**Data representation**

In order to provide a formal description of concepts, terms, and relationships within the WENDI knowledge domain and to make semantic inference possible, we use the Web Ontology Language (OWL) to build the CE ontology and the Resource Description Framework (RDF) in a variety of data interchange formats (e.g. RDF/XML, N3, Turtle, N-Triples) to present CE data based on the CE ontology.

CE OWL ontology is constrained for using in our system: i.e., it is an ontology specific to the datasets used in CE and is not a generalized chemogenomic ontology. Within the ontology we use the following entity classes: Chemical Compound, BioAssay, Journal Article, Gene, and Disease. These entities can be associated by relational ontological terms as shown in Table 1. Also the entity and relational terms can then be combined to express entity-relationship-entity triples, which are suitable for representation in RDF. Some triple examples are given in Table 2.

**Table 1. Examples of Object Properties and Classes for the CE Ontology**

| Properties | Classes | Explanation |
|---|---|---|
| isSimilarTo | Chemical Compound, Chemical Compound | Chemical Compound is similar to Chemical Compound |
| isActiveIn | Chemical Compound, PubChem BioAssay | Chemical is tested active in the bioassay |
| isContainedIn | Chemical Compound, Journal Article | Chemical is contained in the article |
| hasGenes | Pubchem BioAssay/Drug/Journal Article, Gene | Bioassay/Drug/Article has found the gene term related the corresponding text; Bioassay/Drug/Article has a reference to the gene |
| hasDisease | Pubchem BioAssay/Drug/Journal Article, Disease | Bioassay/Drug/Article has found the disease term related the corresponding text; Bioassay/Drug/Article has a reference to the disease |
| isAssociatedWith | Gene, Disease | Gene and disease is associated |
| hasSimilarity | Chemical Compound, Similarity | Chemical has similarity value based on Tanimoto coefficient. |

**Table 2. CE triple examples based on CE Ontology**

| CE Triple Examples |
| --- |
| WO:querycmpd WO:isSimilarTo  WO:cid24871487. |
| WO:cid24871487 rdf:type WO:ChemicalCompound; <br> WO:isActiveIn WO:aid1469. |
| WO:aid1469 rdf:type WO:BioAssay; <br> WO:hasGenes WO:COL4A4. |
| WO:COL4A4 rdf:type WO:Gene; <br> WO:isAssociatedWith WO:Nephritis. |

Figure 2 shows the network of possible relationships representing by above triples expressed in this system. Classes listed in Table 1 are shown in yellow ovals, like Journal Article, Chemical Compound, Gene, Disease, Bioassay, instances are in white ovals, black arrows show direct relationships mined from WENDI, and red arrows show inferred relationships mined from CE, like "Methysergide-Autistic_Disorder-HTR1B", "Methysergide-Lymphoma-CYP1A2" that can be derived from our rule base.

**Figure 2. RDF Network for CE**

**Inference and the Rule Base**

Inference [26], in the context of SWT, is the discovery of new relationships from the known data modelled as a set of (named) relationships between resources and a set of rules automatically. In a mathematical sense, querying is a form of inference (being able to infer some search results from a mass of data, for example) [27]. We make inference to find new inferred relationships between compound and disease.

Once the CE RDF triples are generated, they are loaded into Ont Model Class [28] in Jena [29], a Java Semantic Web Platform. We are performing the rule-based reasoner and forward chaining over RDF graphs. A rule for the rule-based reasoner is defined by a Java Rule object with a list of body terms (premises), a list of head terms (conclusions) and an optional name and optional direction. Each term or ClauseEntry is either a triple pattern, an extended triple pattern or a call to a built-in primitive [29]. Total 8 rules have been defined in the CE system [30], 3 of them with explanations are listed below:

[**Rule 1:** (?QueryCompound WO:isSimiliarTo ?CompoundID),

 (?CompoundID WO:isActiveIn ?Bioassay),

 (?Bioassay WO:isAssociatedWith ?Disease)

-> (?QueryCompound WO:mightHasDisease ?Disease)]

Explanation: A relationship is inferred between a compound and a disease if the query compound is similar to another compound that is active against a PubChem Bioassay, and that Bioassay is associated with a disease.

[**Rule 2:** (?CompoundID WO:isContainedIn ?Journal),

(?Journal WO:hasGene ?Gene),

(?Gene WO:isAssociatedWith ?Disease)

-> (?CompoundID WO:mightHasDisease ?Disease)]

Explanation: A new compound-disease relationship is inferred if there a similar compound and a gene co-occur in a paper abstract, and the gene and disease co-occur in another paper abstract.

[**Rule 3:** (?CompoundID WO:isActiveIn ?Bioassay),

 (?Bioassay WO:hasGene ?Gene),

 (?Gene WO:isAssociatedWith ?Disease)

-> (?CompoundID WO:mightHasDisease ?Disease)]

Explanation: A new compound-disease relationship is inferred if a similar compound is active against a bioassay, the bioassay is associated with a gene, and the gene co-occurs in a paper abstract with the disease.

We selected Methysergide [31] as an example query compound for the following steps. Methysergide is chemically similar to LSD [32], and it antagonizes the effects of serotonin in blood vessels and gastrointestinal smooth muscle, but has few of the properties of other ergot alkaloids.

Table 3 shows three RDF statements of Methysergide taken from CE RDF network. Based on that, we got inferred evidence paths by using above rules. Each statement along with explanation can be found in this Table 3.

**Table 3. CE RDF statement examples**

| RDF Statements | Explanation |
|---|---|
| wo:ctdcid9681 rdf:type wo:ChemicalCompound.<br>wo:querycmpd wo:isSimilarTo wo:ctdcid9681;<br>    wo:hasSimilarity "1.000".<br>wo:ctdcid9681 wo:hasName "Methysergide".<br>wo:HTR1B rdf:type wo:Gene;<br>    wo:isrelatedTo wo:cid9681;<br>    wo:isInferredFrom "pubmedid8743744".<br>wo:Autistic_Disorder rdf:type wo:Disease;<br>    wo:isAssociatedWith wo:HTR1B;<br>    wo:isInferredFrom "pubmedid19038234". | A Methysergide-Autistic Disorder relationship is inferred via rule 2 (gene HTR1B). The similar compound (cid = 9681) is Methysergide itself with similarity = 1, it co-occurs with gene HTR1B in a same paper (pubmed id = 8743744), and HTR1B and Autistic_Disorder are co-occurring in another same paper (pubmed id = 19038234). Then we use rule 2 to establish such relations; |
| wo:ctdcid11865408 rdf:type wo:ChemicalCompound.<br>wo:querycmpd wo:isSimilarTo wo:ctdcid11865408;<br>    wo:hasSimilarity "0.774".<br>wo:ctdcid11865408 wo:hasName "Metergoline".<br>wo:HTR1B rdf:type wo:Gene;<br>    wo:isrelatedTo wo:cid11865408;<br>    wo:isInferredFrom "pubmedid1330643".<br>wo:Autistic_Disorder rdf:type wo:Disease;<br>    wo:isAssociatedWith wo:HTR1B;<br>    wo:isInferredFrom "pubmedid19038234". | A Methysergide-Autistic Disorder relationship is also inferred via rule 2 (again via HTR1B). Although this is the same relationship, a different evidence path considered (we will do path ranking on these evidence paths later); |
| wo:cid5486180 rdf:type wo:ChemicalCompound.<br>wo:querycmpd wo:isSimilarTo wo:cid5486180;<br>    wo:hasSimilarity "0.929".<br>wo:cid5486180 wo:isActiveIn wo:aid410.<br>wo:aid410 rdf:type wo:BioAssay;<br>    wo:hasName "p450-cyp1a2".<br>wo:CYP1A2 rdf:type wo:Gene.<br>wo:aid410 wo:hasGene wo:CYP1A2.<br>wo:CYP1A2 wo:isAssociatedWith wo:Lymphoma.<br>wo:Lymphoma rdf:type wo:Disease. | A Methysergide-Lymphoma relationship is inferred by rule1 (via CYP1A2). |

Methysergide as the query compound, we got a total of 63 evidence paths with different diseases, genes, and journal information. Individual evidence paths can be examined to get to the root data or publications that constitute them. For instance, the Autism link is demonstrated is interesting as the publications identify the link of the compound with HTR1B and the link of HTR1B with Autism. LSD is known to affect the outcome of Autism [33, 34] and thus Methysergide is a reasonable candidate for investigation.

Browsing RDF is clearly difficult, we have thus built an interface that allows the results to be examined and filtered in a user-friendly fashion, more details about the interface shows in the next section. Specifically, evidence paths are clustered by disease, and can be filtered via disease, compounds, assays, genes, gene families, or journal titles. Part of the results for Methysergide are shown in Figure 3, using

"Autistic_Disorder" as the filter, two similar compounds including Methysergide itself are related "Autistic_Disorder" with HTRB1 and a journal article. The results with AID "410" as the filter are shown in Figure 4. Total 20 entries associating with different similar compounds/diseases/genes/references are related the PubChem Bioassay (AID = 410).

**Figure 3. Results related to "Autistic_Disorder" for Methysergide shown in the CE Faced Browser by using Disease filter**

**Figure 4. Results related to AID "410" for Methysergide shown in the CE Faced Browser by using AID filter**

**Path Ranking**

The above process results in often many evidence paths linking compounds and diseases. With a large number of results, we need some way to organize and prioritize these evidence paths. We cluster all the paths based on the different disease terms and then rank the clusters based on the number of evidence paths linking them. Whilst evidence paths are not necessarily fully independent, they do constitute different collections of evidence for the same relationship, and thus strengthen the chances of the relationship being significant.

We employ the following SPARQL query clause to implement this ranking process based on the inferred RDF. It counts the number of properties (?pc) related to each disease first, and then return disease terms (?dis) as descend order on the basis of (?pc).

Select ?dis (count(?p) as ?pc)

WHERE {?dis a wo:Disease;   ?p ?o}

GROUP BY ?dis ORDER BY DESC(?pc)

# Results and discussion

The architecture for CE is shown in Figure 5. CE does data retrieval, data process, and data visualization. When query compound submitted to "Data Controller", a servlet communicating with client and server, Data Controller sends the request to WENDI web service, after that, WENDI XML will be passed to "RDF Model Builder", which handles: CE ontology generation, RDF converting, RDF inference, and path ranking. Ranked paths will be sent back to Data Controller to convert to RDF based JSON file for visualization by using the Faceted Browser. SPARQL query builder, is an additional CE user platform to make customized SPARQL queries based on CE RDF sent back from Data Controller.

**Figure 5. CE Architecture and Path Ranking Flowchart**

### Faceted Browser for CE

CE provides a main web interface, shown in Figure 6. In the Figure, Methysergide is drawn using the JME molecular editor, and its SMILES is transferred to the input box. And the results will be displayed in the Faceted Browser based on an existing tool[35] and allowing multiple filters to be applied.

**Figure 6. CE main Web Interface**

### SPARQL Query Builder for CE

After XML to RDF conversion, CE has RDF triples based on CE ontology. We therefore saw the utility of allowing the direct querying of this RDF data. Since SPARQL is a complex language, we implemented a SPARQL Query Builder to semi-automate this process. The SPARQL query builder for CE is built based on the Sesame triple store [36]. The interface is shown in Figure 7. Starting with a class, the user can add data and object properties associated with it through prompted drop-down boxes. Step by step, the SPARQL query builder provides an intuitive way to translate user question into graph pattern, and then encode it into a SPARQL query.

As an example, given the relationship of Methysergide [31] with HTR1B, LSD and Autism discussed, so we can explore the relationship of similar compounds with the serotonin 5-HT1B receptor (the LSD receptor) with a SPARQL query. We make the SPARQL query in the builder with the following 2 steps to get journal papers including information about "5-HT" receptor:

1) Find similar compounds to Methysergide from the literatures,

Subject: wo:ChemicalCompound

Predict: wo:isContainedIn

Object: wo:journalArticle

2) The titles of the papers should include "5-HT",

Subject: wo: journalArticle

Predict: wo:hasTitle

Object: "5-HT"

The implementation of this query is shown in Figure 7, and the list of journal titles including "5-HT" is shown in the result page of SPARQL Query Builder, in Table 4.

**Figure 7. Main Web Interface of CE SPARQL Query Builder**

**Table 4. List of Journal Titles including "5-HT" receptor**

| Journal Titles |
|---|
| First Pharmacophoric Hypothesis for 5-HT7 Antagonism |
| Novel, Potent, and Selective 5-HT3 Receptor Antagonists Based on the Arylpiperazine Skeleton: Synthesis, Structure, Biological Activity, and Comparative Molecular Field Analysis Studies |
| Synthesis of 2-Piperazinylbenzothiazole and 2-Piperazinylbenzoxazole Derivatives with 5-HT3 Antagonist and 5-HT4 Agonist Properties |
| Novel and Highly Potent 5-HT3 Receptor Agonists Based on a Pyrroloquinoxaline Structure |

**Identification of potential gene targets and diseases for Clozapine**

In order to validate CE, we tested it with well-known drugs as queries, to see how the ordering of the clusters of evidence paths related to known uses and side-effects for these drugs. For example Clozapine [37] has been shown to have superior efficacy when compared to olanzapine [38] in the treatment of schizophrenia [39]. Some known side effects of Clozapine, are cardiomyopathy (deterioration of the function of the myocardium), and cardiac hypertrophy. For this drug, CE indeed predominately returns compound-disease paths that relate to schizophrenia (i.e. schizophrenia has more evidence paths than any other disease). It also identifies side effects of the drug correctly as hypertrophic cardiomyopathy, and cardiovascular system disease, both of which are supported by the literature [40, 41]. This is shown in Figure 8.

**Figure 8. CE results for Clozapine**

**Exploring newly submitted compounds from PubChem**

Pubchem is a popular public database of chemical compounds and their activities against biological assays. Since CE is designed for use with "new" compounds as queries (i.e. compounds for which there is not a lot of data available), we chose a set of very recently-added compounds in PubChem which had no or little associated bioactivity information recorded. This was done using a constrained search in PubChem [42] to return compounds submitted only in 2011.

For example, as shown in Figure 9, the compound with CID 49835692 [43] has no associated bioactivity data recorded. However, through its analysis of similar structures, some significant potential bioactivities and disease associations are suggested by CE.

**Figure 9. More Chemogenomic information for New Compound from PubChem**

We were using RDF network to make inference between compounds and diseases. As the experiments discussed before, not only the most related diseases could be sorted out, but also general guideline will be generated to conduct new compounds analysis. The power of the methodology has been clearly demonstrated to retrieve pertinent information in particular domain without any difficulties engendering by the data tsunami. In addition, it expands the possible usages/linkages within the limited volumes of disease information regarding to a specific compound.

## Conclusions

We present a new approach to the association search of chemical compounds and diseases using semantic inference in this work. Semantic inference produces evidence paths relating compounds and diseases via genes, publication, bioassays and drugs. We previously released an aggregative data-mining tool, WENDI, for drug discovery using aggregate web services. In this paper, we have shown how the application of Semantic Web methods (RDF, SPARQL and OWL ontologies) along with rule-based inference, path ranking and a faceted browse, can produce a tool for exploring new compound-disease associations based on evidence paths from WENDI.

## Future work

The current version of CE explores the chemogenomic information of chemical compounds. In the future, we will consider more efficient ways to mine compound-gene, compound-disease links from more chemogenomic data, and plan to aggregate additional data and inference rules, also increase collaboration with Chem2Bio2RDF in order to enable CE to link with more diverse data. We also intend to expand beyond Chem2Bio2RDF to chemical biology, where we can consider other relations like chemical-gene, chemical-pathway, chemical-side effect, etc. In addition, we would like to add the functionality to process batches of molecules. For this case, we will consider the issues of information summarization and visualization, i.e. how to organize more data in a readable way. Because of the increased volume of data and results, some current algorithms will become out of date. We will also take other ranking algorithms into account such as evidence importance.

## Competing interests

The authors declare that they have no competing interests.

## Authors' contributions

ZQ carried out the majority of the implementation of Chemogenoic Explorer. SYY and ZQ worked on the exploring tools. ZQ and SC worked on the semantic inference. DJW and DY assembled this paper from a previous white paper written by ZQ. All contributed to the intellectual evolution of this project. All authors have read and approved the final manuscript.

## Acknowledgements

This project is supported by Eli Lilly and Company. We thank Bret Davidson for correcting English. We would like to thank the anonymous reviewers for their helpful comments and suggestions.

## References

1. **PubChem** [http://pubchem.ncbi.nlm.nih.gov]

2. **ChEMBL** [http://www.ebi.ac.uk/chembldb/]

3. **ChEBI** [www.ebi.ac.uk/chebi/]

4. **CTD** [http://ctd.mdibl.org]

5. **KEGG** [http://www.genome.jp/kegg/]

6. **Medline** [http://www.nlm.nih.gov/databases/databases_medline.html]

7. **Chemspider** [http://www.chemspider.com/]

8. **LODD** [http://esw.w3.org/HCLSIG/LODD]

9. **Bio2RDF** [http://bio2rdf.org/]

10. Chen, B., Dong, X., Jiao, Dazhi, Wang, H., Zhu, Q., Ding, Y. and Wild, D: **Chem2Bio2RDF: A semantic framework for linking and mining chemogenomic and systems chemical biology data.** *BMC Bioinformatics* 2010, **11**: 255.

11. **PharmGKB** [http://www.pharmgkb.org/]

12. Jesús Bermejo-Muñoz A: **flexible semantic inference methodology to reason about user preferences in knowledge-based recommender systems.** *Knowledge-Based Systems* 2008, **21(4)**:305-320

13. HADACZ, Leo. **Semantic Inference in the Human -Machine Communication.** In: *Lecture Notes in Computer Science.* Berlin-Heidelberg : Springer Verlag, 1999, 353-356.

14. Ruttenberg A., Rees J. A., Samwald M., Marshall M. S. **Life sciences on the semantic web: the Neurocommons and beyond.** *Brief. Bioinform.* **10**, 193–204.

15. **Neurocommons** [http://neurocommons.org/page/Main_Page]

16. Kobayashi, Norio, Ishii, Manabu, Yoshida, Yuko, Makita, Yuko, Matsushima, Akihiro, Mochizuki, Yoshiki, and Toyoda, Tetsuro. **SciNetS Search : Inference Search over an Integrated Life-sciences Database Based on the Semantic Web.** *Nature Precedings* <http://dx.doi.org/10.1038/npre.2010.5083.1> (2010)

17. Xu, Qingwei, Shi, Yixiang, Lu, Qiang, Zhang, Guoqing, Luo, Qingming and Li, Yixue **GORouter: an RDF model for providing semantic query and inference services for Gene Ontology and its associations**. *BMC Bioinformatics*, 2008, **9** (Suppl 1):S6

18. Zhu, Q., Lajiness, M.S., Ding, Y., Wild, D.J. **WENDI: A tool for finding non-obvious relationships between compounds and biological properties, genes, diseases and scholarly publications.** *Journal of Cheminformatics*, 2010, **2**:6

19. PubChem3D is a 3D version of PubChem, in which we have generated a single conformer for 99% of PubChem using the smi23d suite of programs.

20. **DrugBank** [http://www.drugbank.ca]

21. **MRTD** [http://www.fda.gov/AboutFDA/CentersOffices/CDER/ucm092199.htm]

22. **HuGE Navigator** [http://hugenavigator.net]

23. **Phenopred Matrix** [www.phenopred.org]

24. **Gene Ontology** [www.geneontology.org]

25. **XML DOM** [http://www.w3schools.com/xml/xml_dom.asp]

26. **Semantic Inference** [http://www.w3.org/standards/semanticweb/inference.html]

27. **The Semantic Web: An Introduction** [http://infomesh.net/2001/swintro/]

28. **OntModel** [http://jena.sourceforge.net/javadoc/com/hp/hpl/jena/ontology/OntModel.html]

29. **Jena** [http://jena.sourceforge.net/]

30. **Chemogenmic Explorer Rules**

    [https://cheminfov.informatics.indiana.edu:8443/chemogenomic_explorer/Rules.doc]

31. **Methysergide**

    [http://pubchem.ncbi.nlm.nih.gov/summary/summary.cgi?cid=9681]

32. **LSD** [http://en.wikipedia.org/wiki/Lysergic_acid_diethylamide]

33. BENDER L, GOLDSCHMIDT L, SANKAR DV. **Treatment of autistic schizophrenic children with LSD-25 and UML-491.** *Recent Adv Biol Psychiatry* 1961, **4**:170–179.

34. Simmons JQ, Benor D, Daniel D. **The variable effects of LSD-25 on the behavior of a heterogeneous group of childhood schizophrenics.** *Behavioral Neuropsychiatry* 1972, **3**:10–24.

35. **Lonewell** [http://simile.mit.edu/wiki/Longwell]

36. **Sesame** [http://www.openrdf.org/index.jsp]

37. **Clozapine** [http://en.wikipedia.org/wiki/Clozapine#Withdrawal_effects]

38. **Olanzapine** http://en.wikipedia.org/wiki/Olanzapine

39. Volavka J. **Clozapine may be more effective than olanzapine for reducing suicidal behaviour in people with schizophrenia at high risk.** *Evid. Based Ment. Health* 2003,**6**:93-93.

40. Pastor CA, Mehta M. **Masked clozapine-induced cardiomyopathy.** *J Am Board Fam Med* 2008, **21**:70–74

41. Killian JG, Kerr K, Lawrence C, Celermajer DS. **Myocarditis and cardiomyopathy associated with clozapine.** *Lancet* 1999, **354**:1841–1845

42. **PubChem advanced search**

    [http://www.ncbi.nlm.nih.gov/sites/entrez?db=pccompound&TabCmd=Limits]

43. **PubChem CID = 49835692**

    [http://pubchem.ncbi.nlm.nih.gov/summary/summary.cgi?cid=49835692&loc=ec_rcs]

# Figures

**Figure 1 - WENDI Architecture**
WENDI main web interface is show at the upper right corner

**Figure 2 - RDF Network for CE**

**Figure 3 - Results related to "Autistic_Disorder" for Methysergide shown in the CE Faced Browser by using Disease filter**

**Figure 4 - Results related to AID "410" for Methysergide shown in the CE Faced Browser by using AID filter**

**Figure 5 - CE Architecture and Path Ranking Flowchart**

**Figure 6 - CE main Web Interface**

**Figure 7 - Main Web Interface of CE SPARQL Query Builder**

**Figure 8 - CE results for Clozapine**

**Figure 9 - More Chemogenomic information for New Compound from PubChem**

# Tables

**Table 1 - Examples of Object Properties and Classes for the CE Ontology**

**Table 2 - CE triple examples based on CE Ontology**

**Table 3 - CE RDF statement examples**

**Table 4 - List of Journal Titles including "5-HT" receptor**

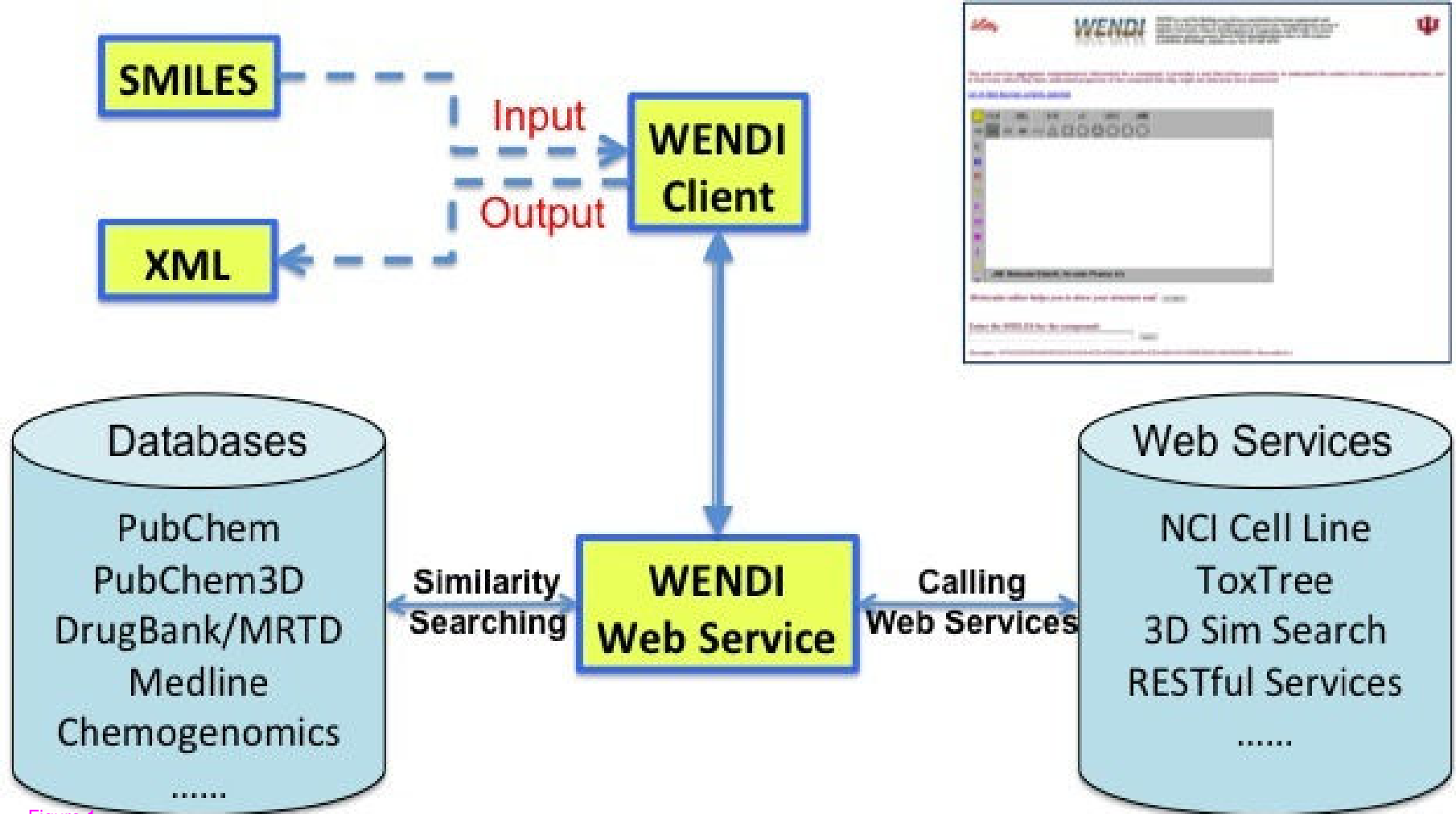

Figure 1

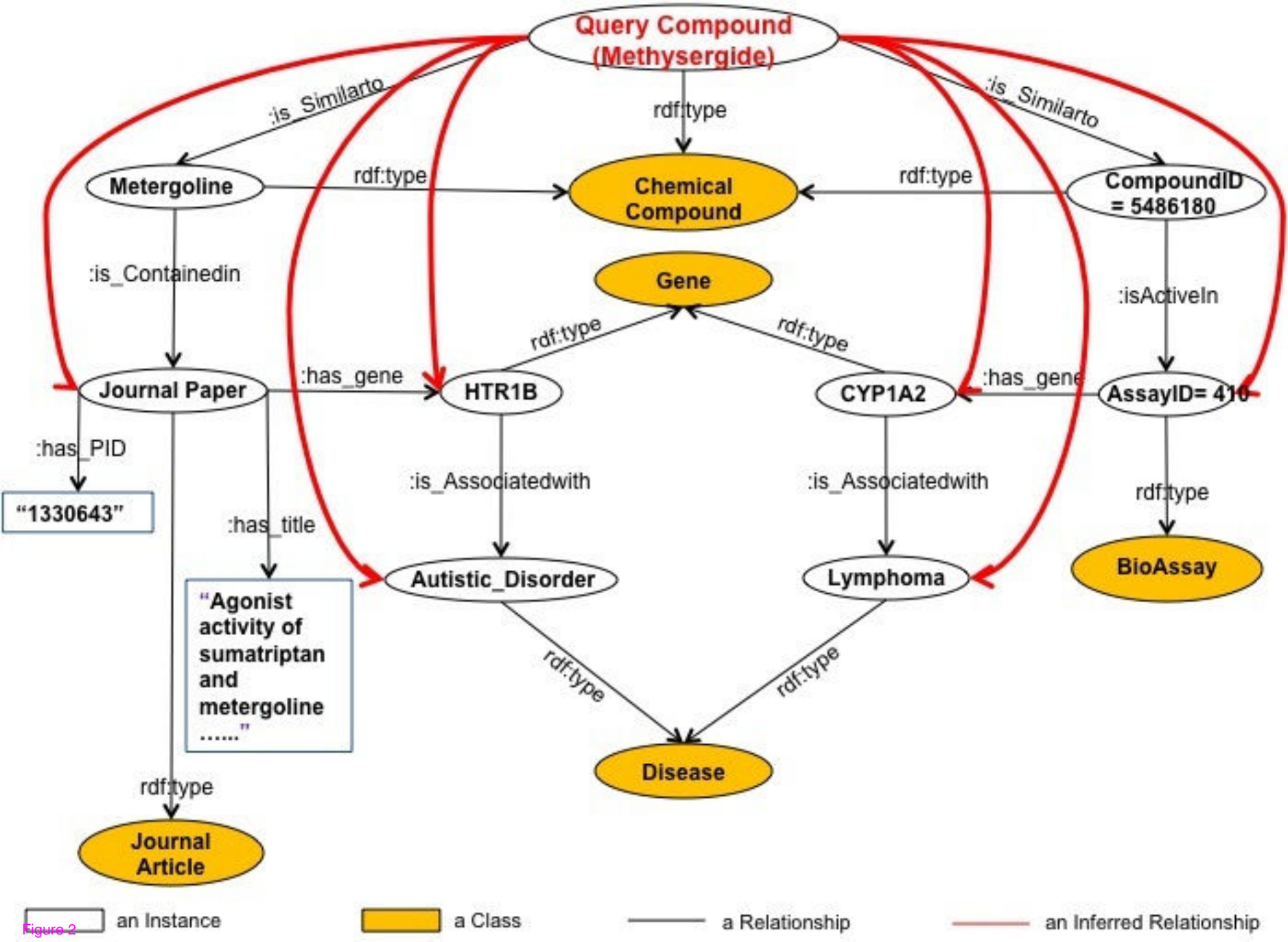

Figure 2

# Result for CCC(CO)NC(=O)C1CN(C2CC3=CN(C4=CC=CC(=C34)C2=C1)C)C

Back to Chemogenomic Explorer Main Page: https://cheminfov.informatics.indiana.edu:8443/chemogenomic_explorer/index.jsp

**2** Autistic_Disorder filtered from 63 originally (Reset All Filters)

sorted by: rank and type; then by... • ☑ grouped as sorted

## 11 (2)

### Autistic_Disorder (2)

1.

| Disease | CID | Structure | AID | GENE | GENE_Family | Journal |
|---|---|---|---|---|---|---|
| Autistic_Disorder | 9681 | | Not available | HTR1B | 5HTR | 19038234 |

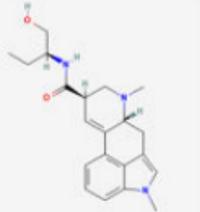

2.

| Disease | CID | Structure | AID | GENE | GENE_Family | Journal |
|---|---|---|---|---|---|---|
| Autistic_Disorder | 11865408 | | Not available | HTR1B | 5HTR | 19038234 |

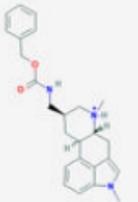

Print Result

**Disease**
- Urogenital_Abnormalities
- 2 Autistic_Disorder ☑
- 2 Congenital_Abnormality
- 2 deformity
- 2 endocrine_system_disease
- Female_Reproductive_System_Disord

**CID**
- 1 11865408
- 1 9681

**AID**
- 2 Not available

**Gene**
- 2 HTR1B

**Gene_Family**
- 2 5HTR

**Jornal**
- 2 19038234

Figure 3

# Result for CCC(CO)NC(=O)C1CN(C2CC3=CN(C4=CC=CC(=C34)C2=C1)C)C

**20** Items filtered from 63 originally (Reset All Filters)

sorted by: rank and type; then by... • ☑ grouped as sorted

Print Result

**Disease**
- 2 Congenital_Abnormality
- 2 deformity
- 2 Disorder_of_fluid_AND_OR_electrol
- 2 Electrolyte_imbalance
- 2 Endocrine_Syndrome

## 03 (2)

### Disorder_of_fluid_AND_OR_electrolyte (2)

1.

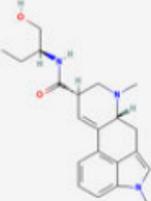

| Disease | CID | Structure | AID | GENE | GENE_Family | Journal |
|---|---|---|---|---|---|---|
| Disorder_of_fluid_AND_OR_electrolyte | 6604697 | | 410 | CYP1A2 | CYP | Not available |

**CID**
- 10 5486180
- 10 6604697

**AID**
- 20 **410**

**Gene**
- 20 CYP1A2

**Gene_Family**
- 20 CYP

**Jornal**
- 20 Not available

2.

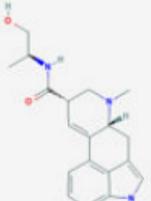

| Disease | CID | Structure | AID | GENE | GENE_Family | Journal |
|---|---|---|---|---|---|---|
| Disorder_of_fluid_AND_OR_electrolyte | 5486180 | | 410 | CYP1A2 | CYP | Not available |

## 04 (2)

Figure 4

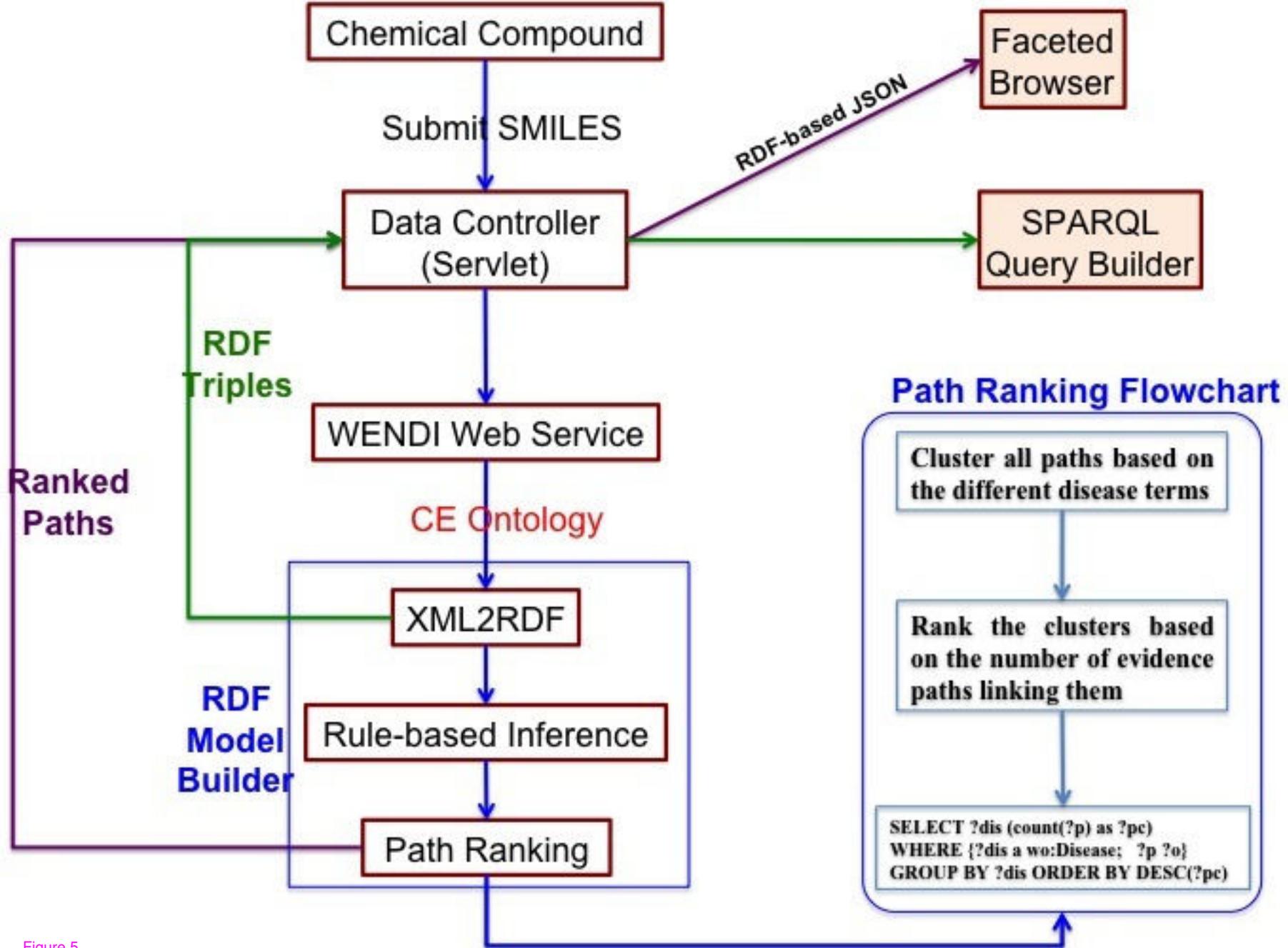

Figure 5

# Chemogenomic Explorer

Chemogenomic Explorer automatically gathers evidence relating compounds to genes and diseases using output from WENDI, which is the product of a collaboration between the cheminformatics group at Indiana University, School of Informatics & Computing, and Eli Lilly & Company.

Molecular editor helps you to draw your structure and [Get SMILES]

**Enter SMILES for the compound:**

CCC(CO)NC(=O)C4C=C3c1ccccc2c1c(cn2C)CC3N(C)C4 [Submit]

(Example: CCC(CO)CC(O1)OC2CC(CC3=C(C4=C(C(=C23)O)C(=O)C5=C(C4=O)C=CC=C5OC)O)(C(=O)CO)O)N)O ( Doxorubicin ))

| SPARQL Query Builder [BACK] | | |
|---|---|---|
| Subject | Predicate | Object |
| wo:ChemicalCompound | wo:isContainedIn [Delete] | wo:JournalArticle |
| [Add Property] | | |
| wo:JournalArticle [Delete] | wo:hasTitle [Delete] | 5-HT |
| [Add Property] | | |

[Generate Query]

```
PREFIX rdfs: <http://www.w3.org/2000/01/rdf-schema#>
PREFIX rdf: <http://www.w3.org/1999/02/22-rdf-syntax-ns#>
PREFIX wo: <http://www.cheminfo.informatics.indiana.edu/WendiOntology.owl#>

SELECT distinct ?JournalArticle1_hasTitle
WHERE{
?ChemicalCompound1 rdf:type wo:ChemicalCompound .
?ChemicalCompound1 wo:isContainedIn ?JournalArticle1 .
?JournalArticle1 rdf:type wo:JournalArticle .
?JournalArticle1 wo:hasTitle ?JournalArticle1_hasTitle .
FILTER REGEX (str(?JournalArticle1_hasTitle), '5-HT', 'i')}
```

[Query]

Figure 7

**Hypertrophic_Cardiomyopathy (4)**

1.

| Disease | CID | Structure | AID | GENE | GENE_Family | Journal |
|---|---|---|---|---|---|---|
| Hypertrophic_Cardiomyopathy | 2820 | 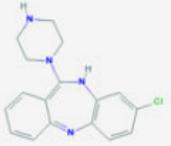 | 410 | CYP1A2 | CYP | Not available |

2.

| Disease | CID | Structure | AID | GENE | GENE_Family | Journal |
|---|---|---|---|---|---|---|
| Hypertrophic_Cardiomyopathy | 3233486 | 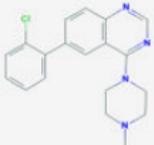 | 410 | CYP1A2 | CYP | Not available |

3.

| Disease | CID | Structure | AID | GENE | GENE_Family | Journal |
|---|---|---|---|---|---|---|
| Hypertrophic_Cardiomyopathy | 2818 | 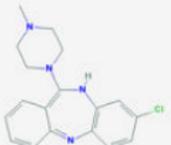 | 410 | CYP1A2 | CYP | Not available |

4.

| Disease | CID | Structure | AID | GENE | GENE_Family | Journal |
|---|---|---|---|---|---|---|
| Hypertrophic_Cardiomyopathy | 3233486 | 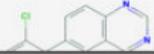 | 899 | CYP2C19 | CYP | Not available |

Figure 8

---

20 cancer
20 Hepatitis
12 syndrome

**CID**
2 3233486
1 2818
1 2820

**AID**
3 410
1 899

**Gene**
3 CYP1A2
1 CYP2C19

**Gene_Family**
4 CYP

**Jornal**
4 Not available

# Result for CC1(COC(C(C1NC)O)OC2C(CC(C(C2O)OC3C(CC=C(O3)CNCC4CCNCC4)N)N)NC(=O)C(CN)O)O

8 Items

sorted by: rank and type; then by... • ☑ grouped as sorted

1.

| Disease | CID | Structure | AID | GENE | GENE_Family | Journal |
|---|---|---|---|---|---|---|
| Abnormalities_Drug-Induced | 31703 | 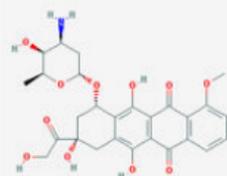 | Not available | EPHX1 | Not Available | 2336087 |

2.

| Disease | CID | Structure | AID | GENE | GENE_Family | Journal |
|---|---|---|---|---|---|---|
| Adenocarcinoma | 3033521 | 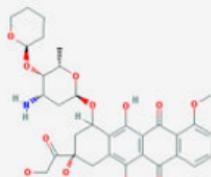 | Not available | PGR | Not Available | 15639718 |

3.

| Disease | CID | Structure | AID | GENE | GENE_Family | Journal |
|---|---|---|---|---|---|---|
| Adenomatous_Polyposis_Coli | 23724849 | 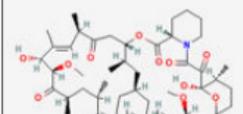 | Not available | AKT1 | Ion Channel Families | 17942926 |

**Print Result**

**Disease**
1. Abnormalities_Drug-Induced
1. Adenocarcinoma
1. Adenomatous_Polyposis_Coli
   Attention_Deficit_Disorder_with_Hyperactivity
1. Autistic_Disorder
1. Breast_Neoplasms
1. Colitis_Ulcerative
1. HIV_Seropositivity

**CID**
1. 23724849
1. 2826713
1. 3033521
1. 31703
1. 5459119
1. 6048582
1. 65907
1. 9576002

**AID**
8. Not available

**Gene**
1. AKT1
   ALB
   CHRNA7
   CYP3A4
   EPHX1

Figure 9